\documentclass[aps,prl,twocolumn,superscriptaddress,showpacs,english]{revtex4-1}
\usepackage[T1]{fontenc}
\usepackage[latin9]{inputenc}
\usepackage{babel}
\usepackage{amsmath}
\usepackage{amssymb}
\usepackage{wasysym}
\usepackage{graphicx}
\usepackage{xcolor}
\definecolor{mydarkgreen}{rgb}{0.0,0.5,0.0}
\definecolor{darkgreen2}{rgb}{0.0,0.3,0.0}
\definecolor{darkgreen2}{rgb}{0.0,0.0,0.0}
\usepackage[linktocpage=true,
  colorlinks=true, 
  pdfborder={0 0 0},
  linkcolor=blue,
  citecolor=darkgreen2,
  filecolor=yellow,
  urlcolor=mydarkgreen,
  bookmarks,
  pdftitle={paper},
  pdfauthor={},
]{hyperref}

\makeatletter

\usepackage[caption=false]{subfig}
\usepackage{afterpage}
\usepackage{babel}

\@ifundefined{showcaptionsetup}{}{%
 \PassOptionsToPackage{caption=false}{subfig}}
\usepackage{subfig}
\makeatother
\newcommand{\zse}{z$_{\text{Se}}$}

\begin{document}
\global\long\def\figurename{Fig.}

\global\long\def\hitc{hi-$T_{\text{c}}$}
\global\long\def\tc{$T_{\text{c}}$}

\title{Ab-initio theory of Iron based superconductors}

\author{F. Essenberger} 
\affiliation{Max Planck Institute of Microstructure Physics, Weinberg 2, D-06120 Halle, Germany.}
\author{A. Sanna}
\affiliation{Max Planck Institute of Microstructure Physics, Weinberg 2, D-06120 Halle, Germany.}
\author{P. Buczek}
\affiliation{Max Planck Institute of Microstructure Physics, Weinberg 2, D-06120 Halle, Germany.}
\author{A. Ernst}
\affiliation{Max Planck Institute of Microstructure Physics, Weinberg 2, D-06120 Halle, Germany.}
\affiliation{Wilhelm-Ostwald-Institut f\"ur Physikalische und Theoretische Chemie,  Universit\"at Leipzig, Linn\'estra{\ss}e 2, 04103 Leipzig, Germany}
\author{L. Sandratskii} 
\affiliation{Max Planck Institute of Microstructure Physics, Weinberg 2, D-06120 Halle, Germany.}
\author{E.K.U. Gross}
\affiliation{Max Planck Institute of Microstructure Physics, Weinberg 2, D-06120 Halle, Germany.}

\date{\today}
\begin{abstract}
We report the first-principles study of superconducting 
critical temperature and superconducting properties of Fe-based superconductors
taking into account on the same footing phonon, charge and spin-fluctuation mediated Cooper pairing.
We show that in FeSe this leads to a modulated s$\pm$ gap symmetry, and that the 
 antiferromagnetic paramagnons are the leading mechanism for superconductivity in FeSe,
overcoming the strong repulsive effect of both phonons and charge pairing. 

\end{abstract}
\maketitle
The discovery of the iron based superconductors (FeSC) in the last
decade\citep{Nature.453.376,PhysRevLett.101.107006,PhysRevLett.102.177005,Nature.464.183}
has been a crucial event in the history of superconductivity (SC).
Showing that high temperature (\hitc) SC is not a unique property of  
the cuprates\citep{ZeitschriftPhysB.64.189}, and that it could
occur in remarkably different classes of systems\citep{RevModPhys.78.17,RevModPhys.60.585,RevModPhys.83.1589,NatPhys.6.645,Physics.2.60}.
This discovery renewed the hope to find a room temperature superconductor, probably the most desired system in solid state physics.

In order to facilitate the search for new \hitc\ materials,
it is  highly desirable to achieve a theoretical understanding of the physical mechanism 
of \hitc\ SC. On a microscopic level, a SC state is created by the
pairing of electrons to form Cooper pairs~\citep{PhysRev.108.1175,nature.2.134}.
Hence, the essential question that theorists try to answer
is what causes this attractive pairing interaction.

A coupling provided by phonons has been ruled out quickly after the
discovery of Fe based superconductors because the electron-phonon (el-ph) coupling is 
 much too weak to explain the observed high transition temperatures\citep{PhysRev.125.1263,PhysRevB.86.064437}
and no clear isotope effect has been measured
\citep{PhysRevB.44.5322,PhysRevLett.58.2337,PhysRevLett.103.257003,PhysRevLett.105.037004,nature.459.64}.
A large number of different theories have been proposed:
resonating valence bond \citep{Science.235.4793,PhysRevLett.58.2790},
fluctuation exchange\citep{PhysRevB.82.024508,PhysRevB.79.220502},
functional renormalization group\citep{PhysRevB.84.235121,RevModPhys.66.129},
orbital fluctuations\citep{PhysRevB.82.144510,PhysRevLett.104.157001},
charge-fluctuations\citep{PhysRevB.35.8869,PhysRevB.44.12500,PhysRevB.84.140505,PhysRevLett.111.057006,PhysRevLett.111.237001},
spin-fluctuations\citep{nature.5.141,Manske,Rep.Prog.Phys.74.124508}
(SF). The scientific community remains far from a general consensus
on which is the dominant coupling mechanism.

Among the different theories the ones based on magnetism are, in our
opinion, the most promising ones, since in both cuprates and FeSC
 the superconductivity appears close to an antiferromagnetic
(AFM) phase\citep{RevModPhys.83.1589,RevModPhys.60.585}.
 Approaching the AFM phase, if the transition is of second order, the magnetic susceptibility
 will become large, and eventually diverge. 
This implies that spin-fluctuations could become strong enough
to overcompensate for the direct electron-electron repulsion and trigger
the SC condensation. This idea is supported by strong experimental\citep{PhysRevLett.102.177005}
and theoretical\citep{RevModPhys.84.1383} arguments.

SF mediated pairing has been extensively investigated in the realm
of Hubbard like models\cite{HM1,HM2,HM3,HM4,Rep.Prog.Phys.74.124508}.
Since this approach necessarily involves a set of parameters,
it does not allow genuine predictions of 
the critical temperature. Therefore, albeit very useful as a tool for 
 a general physical understanding, it does not directly help in the search 
for new superconducting systems with desired 
properties.

The only way for theory to take the lead in the search for new and
better \hitc\ materials is to develop a quantitatively 
predictive \emph{ab-initio} theory and solve the dilemma of the pairing mechanism.

In this work we take a step in this direction by constructing a many-body
perturbation based effective interaction, solely from first-principles calculations,
and use it within density functional theory for superconductors (SCDFT)
\citep{PhysRevLett.60.2430,PhysRevB.72.024545,PhDLueder,PhdMarques}.
While SCDFT proved to be highly  reliable in the description of
el-ph superconductors\citep{PhysRevB.85.184514,PhysRevB.79.104503,PhysRevB.75.054508,IOPscience.22.034006,PhysRevB.75.020511,PhysRevLett.96.047003,PhysRevLett.111.057006},
its spin-fluctuation extension has been tested, until now, just
on a simple electron gas model\citep{our_SF_scdft}. As neither
the functional, nor the theoretical framework, contain any adjustable parameters,
this scheme promises the prediction of \tc, symmetry of the order
parameter and excitation spectrum, just from knowledge of the chemical
structure of the material.

In the present work we apply the scheme to the FeSC. We choose this
family because the metallic parent state can be better described than
the Mott-insulator state in the cuprates. 
However we will observe that an  unsatisfactory description of the parent compound affects
the predictive power of our theoretical approach.

The first conceptual step in the SCDFT scheme currently in use is the assumption
of a second order phase transition between the SC phase and its metallic
parent compound. 
This implies that \tc\ can be estimated
by taking the electronic structure of the metallic non-superconducting phase
as a starting
point to act-on with a pairing field computed from first principles.
So, the very starting point of the
theory is an approximation for the quasi-particle states of the parent
metal, that we do by taking the DFT Kohn-Sham (KS) band structure \citep{PhysRev.140.A1133,PhysRev.136.B864}.

\begin{figure}
\includegraphics[width=\columnwidth]{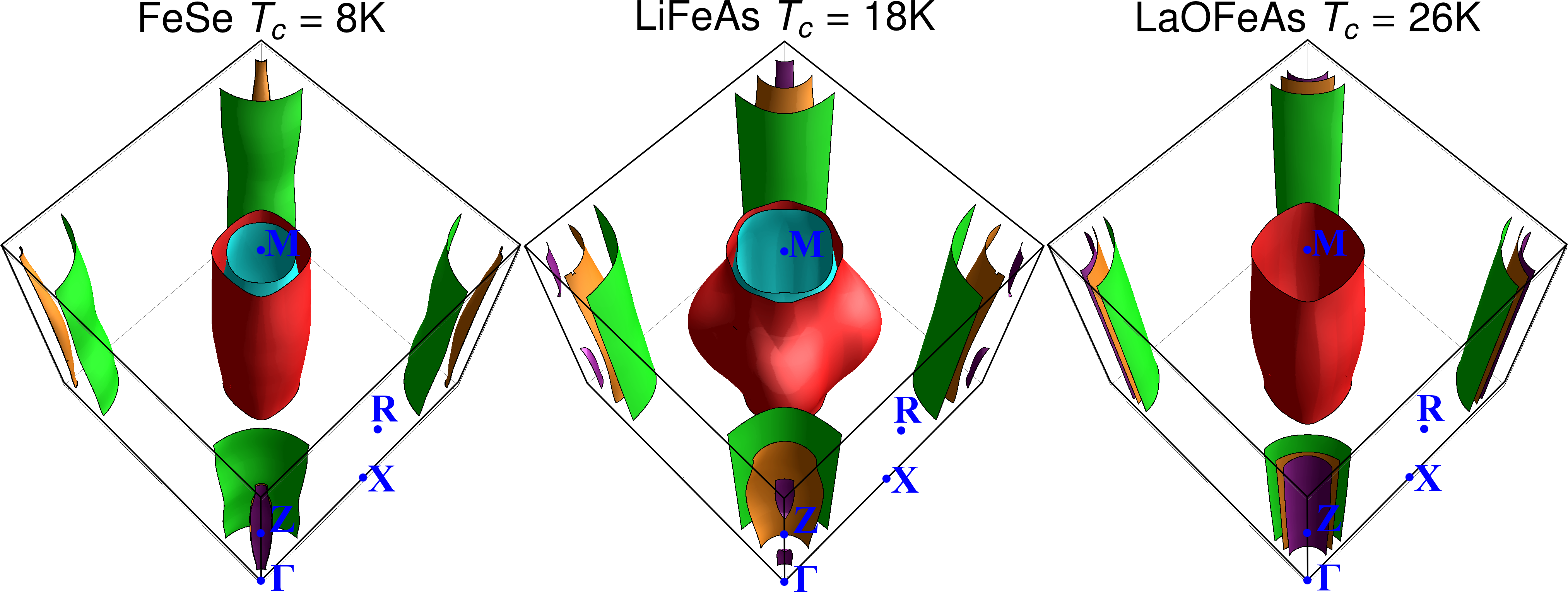}
\caption{(color online) Fermi surface for three Iron superconductors in the tetragonal unit cell.\label{fig:Fermi}}
\end{figure}

The calculated Fermi surface (FS) for FeSe, LiFeAS and LaOFeAs (representatives
of the 11, 111 and 1111 family of FeSC) are shown in Fig.~\ref{fig:Fermi}.
All systems feature the same characteristics: Hole FS around the $\Gamma$-point
forming a barrel and electron FM around the $M$-point which may
not extend along the whole $k_{z}$ direction forming the pockets.
The calculations have been done using a state of the art plane wave
code \citep{QE-2009} and have been cross-checked with an all-electron
linearized augmented plane wave code \citep{ElkCode}. Since the results
are very sensitive to the atomic positions a full lattice relaxation
is performed \citep{PhysRevB.86.060412,PhysRevB.78.085104}. The calculated FS 
are in reasonable agreement with the results of the ARPES measurements \citep{NatPhys.6.645,RevModPhys.83.1589}.

However, the pronounced nesting between the two parts of the FS with
a nesting vector of $\mathbf{q}_{{\scriptscriptstyle \text{M}}}=\left(\frac{\pi}{a},\frac{\pi}{a},0\right)$
leads to an instability  with respect to the formation of the stripe ordered AFM state. 
All compounds converge to an AFM ground state with a large moment of about $1\ \mu_{\text{B}}$ per iron atom. This is
in striking contrast with the experimental observation \citep{NatureMat.10.932},
and a well known problem of DFT calculations for this class of systems \citep{PhysRevB.78.085104,nature.5.141}.
It has been suggested that the reason for this disagreement
is that conventional DFT functionals like LSDA or GGA do not account for
  dynamic zero-point spin  fluctuations that are strong here.
This leads to the overestimation of the stability of the long range magnetic order \citep{NatureMat.10.932}.

In the constraction of the SCDFT SF functional {\citep{our_SF_scdft}} 
an important role is played by the spin susceptibility $\chi_{zz}$.  
The standard calculation of this quantity cannot be performed 
 if the system has an artificial instability with respect to the
formation of a long range magnetic order.

{\color{darkgreen2} The solution of this problem could come from the use an improved
DFT functional perhaps going beyond standard Kohn-Sham scheme. 
However, such a calculational scheme is not yet developed within ab-initio methods, 
and we have to opt for another solution. We will consider two different 
ways to deal with the problem that will help us to examine the
dependence of the superconducting transition temperature on the
details of the accepted approach.}
The first method we consider is to compute the magnetic susceptibility
by scaling down the exchange-correlation (xc) field, as has been proposed
 in Ref.~\onlinecite{PhysRevB.86.064437}.
By introducing a scaling parameter $\alpha$ the magnetic response
function can be written as\cite{PhysRevLett.52.997,GrossKohnLinearResponse}:

\begin{equation}\label{eq:chi}
\chi_{zz}\left(\mathbf{q}\omega\right)=\frac{\chi^{{\scriptscriptstyle \text{KS}}}\left(\mathbf{q}\omega\right)}{1-\alpha f_{\text{xc}}\left(\mathbf{q}\omega\right)\chi^{{\scriptscriptstyle \text{KS}}}\left(\mathbf{q}\omega\right)}.
\end{equation}

By using a sufficiently small $\alpha$ ($<1$) one avoids the singularity of the susceptibility 
corresponding to the phase transition to the magnetic ordered state.
Instead the susceptibility features finite-hight peaks corresponding
to paramagnons.
The energy, lifetime and intensity
of the paramagnons depend on the value of $\alpha$ and all materials
feature a critical value $\alpha_{\text{c}}$ for which the 
susceptibility
$\chi_{zz}\left(\mathbf{q}_{{\scriptscriptstyle \text{M}}}\right)$
diverges.

\begin{figure}
\includegraphics[width=1\columnwidth]{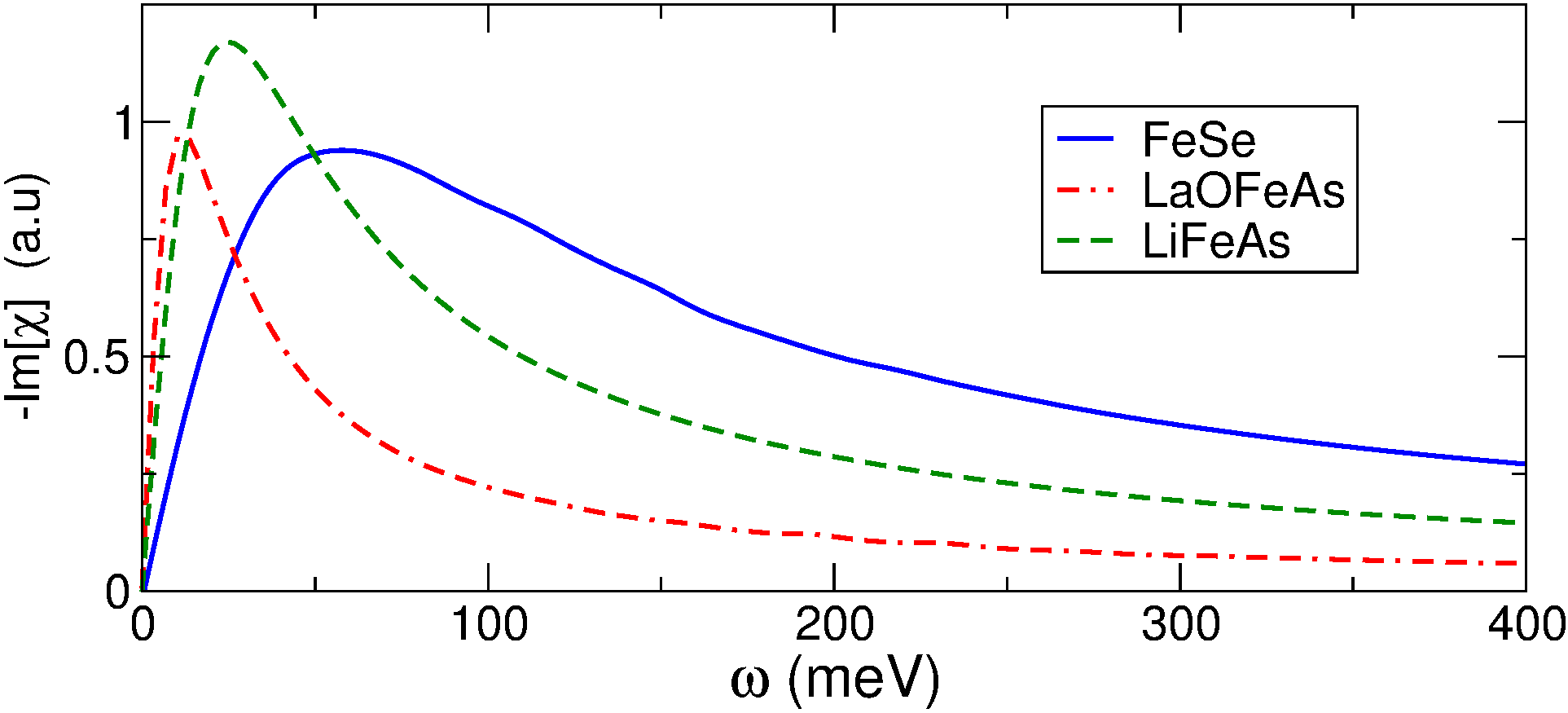}\vspace{-3mm}
\caption{(color online) Spin scuceptibilities 
$\text{Im}[\chi_{zz}\left(\mathbf{q}={\text{M}},\omega,\mathbf{G}=0,\mathbf{G}'=0\right)]$ 
for LiFeAs, LaOFeAs and FeSe, computed at $\frac{\alpha}{\alpha_c}=0.95$.}
\label{fig:ResponseOverview} 
\end{figure}

In Fig.~\ref{fig:ResponseOverview} we show the Im$[\chi_{zz}\left(\mathbf{q}_{{\scriptscriptstyle \text{M}}}\right)]$
of representatives of the 11,111 and 1111 families 
using $\frac{\alpha}{\alpha_{\text{c}}}=0.95$.
All three compounds show a peak in the low-energy region
featuring the presence of paramagnon-type fluctuations.
In the rest of the paper we will focus on one of them: FeSe. 

The SCDFT theoretical framework taking into account spin-fluctuation
effects has been discussed in Ref.~\onlinecite{our_SF_scdft}.
This approach considers  SF, el-ph and Coulomb (charge) pairing
on the same footing. {\color{darkgreen2} First, to keep the formalism complexity at minimum,
we consider a multiband isotropic approximation, meaning that states
and pairings are averaged over Brillouin zone volumes on isoenergy surfaces.
For example, the Kohn-Sham potential\cite{PhysRevB.72.024545} of the SCDFT system $\Delta_{n{\bf k}}$ is averaged as:
\begin{equation}
\Delta_{n\mathbf{k}}\approx\Delta_{n}\left(E\right):=\sum_{\mathbf{k}}\frac{\delta\left(\epsilon_{n\mathbf{k}}-E\right)}{N_{n}\left(E\right)}\Delta_{n\mathbf{k}}\label{eq:isotropeApp}
\end{equation}
This way we can group together the holes (see Fig.~\ref{fig:Fermi})
around the $\Gamma-$point (labeled with $n=+$) and the electrons
at the $M$-point (labeled with $n=-$).} Within this approximation
only $s$-wave pairing symmetries are possible, 
including the $s\pm$ symmetry suggested in Ref.~\onlinecite{Nature.464.183}. 
We will release this averaging approximation at the end of the paper.

Under this approximation we account for: 1) the phononic interaction
by a band-resolved Eliashberg function $\alpha^{2}F_{ij}$ (detailed
definitions can be found in Ref.~\onlinecite{PhysRevB.85.184514}
and references therein); 2) the Coulomb repulsion by the average of
the RPA screened matrix elements (as it was already done in several
previous applications of SCDFT, more details can be found in Ref.~\onlinecite{PhysRevB.75.020511,IOPscience.22.034006};
3) the SF term containing an effective interaction mediated by paramagnon
excitations $\Lambda^{{\scriptscriptstyle \text{SF}}}$ as introduced
in Ref.~\onlinecite{our_SF_scdft}, and averaged as $\Lambda_{n\mathbf{k}n'\mathbf{k}'}\left(\omega\right)\longrightarrow\Lambda_{{\scriptscriptstyle ij}}\left(E,E',\omega\right)\text{ with }i,j\in\left\{ +,-\right\} $.
The structure of phononic, SF and Coulomb contributions is very different in nature. 
For the Coulomb interaction we use a static screening
that proved to be highly reliable in phononic SCDFT\citep{PhysRevB.72.024546,PhysRevLett.96.047003,PhysRevB.75.020511,Floris200745,PhysRevLett.94.037004}.
{\color{darkgreen2} The SF and phonon contributions, on the other hand, have a strong
structure in $\omega$ with a low energy characteristic frequency. Therefore
it is essential to take frequency dependence into account. 
In particular these interactions are negligible at energy scales large with respect to this characteristic one.}
 This allows to safely disregard the $E$ dependence of the interaction
focusing on its behavior near the Fermi level \citep{Allen19831,PhysRev.117.648}.
These considerations then translate into the following approximation
scheme: 
\begin{align*}
\alpha^{2}F_{{\scriptscriptstyle ij}}\left(E,E',\omega\right) & \approx\alpha^{2}F_{{\scriptscriptstyle ij}}\left(\epsilon_{{\scriptscriptstyle \text{F}}},\epsilon_{{\scriptscriptstyle \text{F}}},\omega\right)\text{ (attractive)}\\
w_{{\scriptscriptstyle ij}}\left(E,E',\omega\right) & \approx w_{{\scriptscriptstyle ij}}\left(E,E',0\right)\text{ (repulsive)}\\
\Lambda_{{\scriptscriptstyle ij}}^{{\scriptscriptstyle \text{SF}}}\left(E,E',\omega\right) & \approx\Lambda_{{\scriptscriptstyle ij}}^{{\scriptscriptstyle \text{SF}}}\left(\epsilon_{{\scriptscriptstyle \text{F}}},\epsilon_{{\scriptscriptstyle \text{F}}},\omega\right)\text{ (repulsive)}
\end{align*}

\begin{figure}
\begin{centering}
\begin{minipage}{1.0\columnwidth}
\hspace{0.02\columnwidth}\includegraphics[width=0.82\columnwidth]{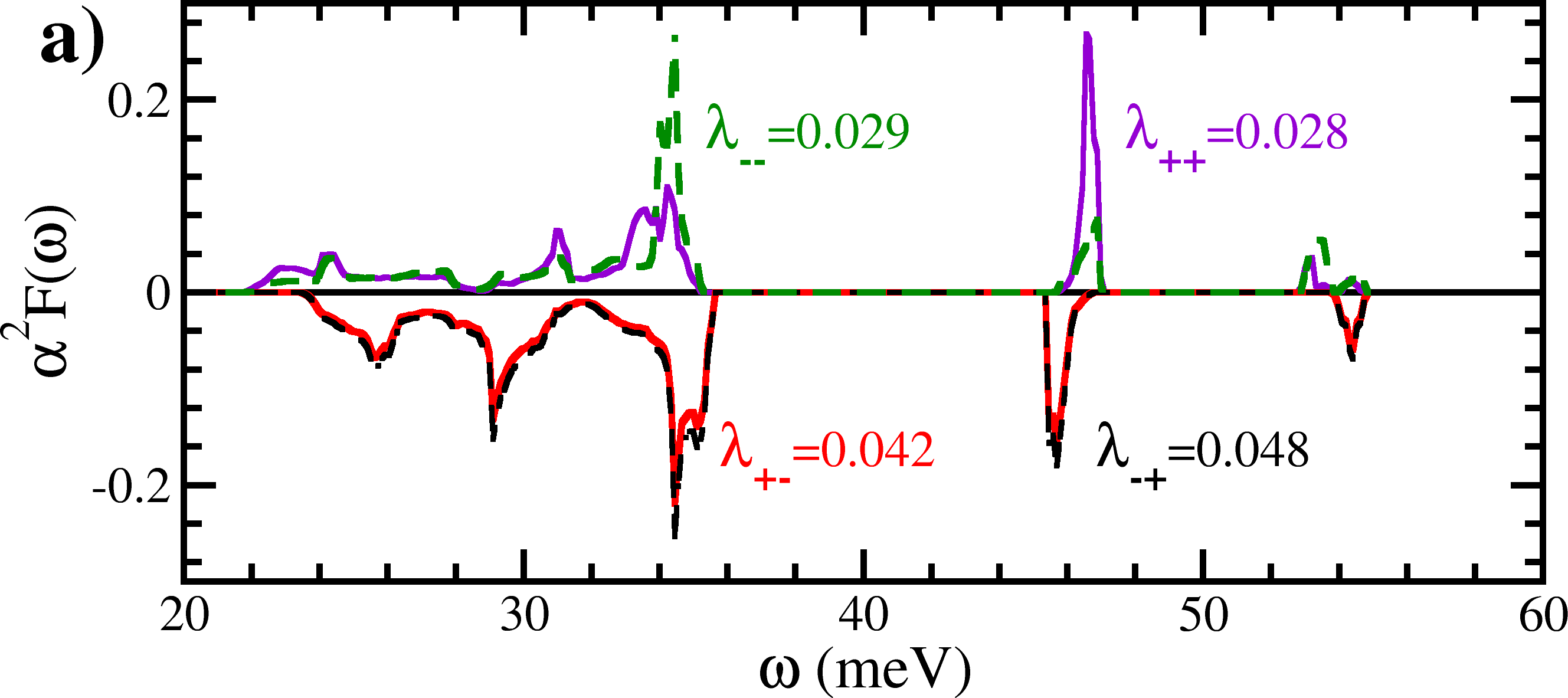}
\end{minipage}
\begin{minipage}{1.0\columnwidth}
\hspace{0.1\columnwidth}\includegraphics[width=0.85\columnwidth]{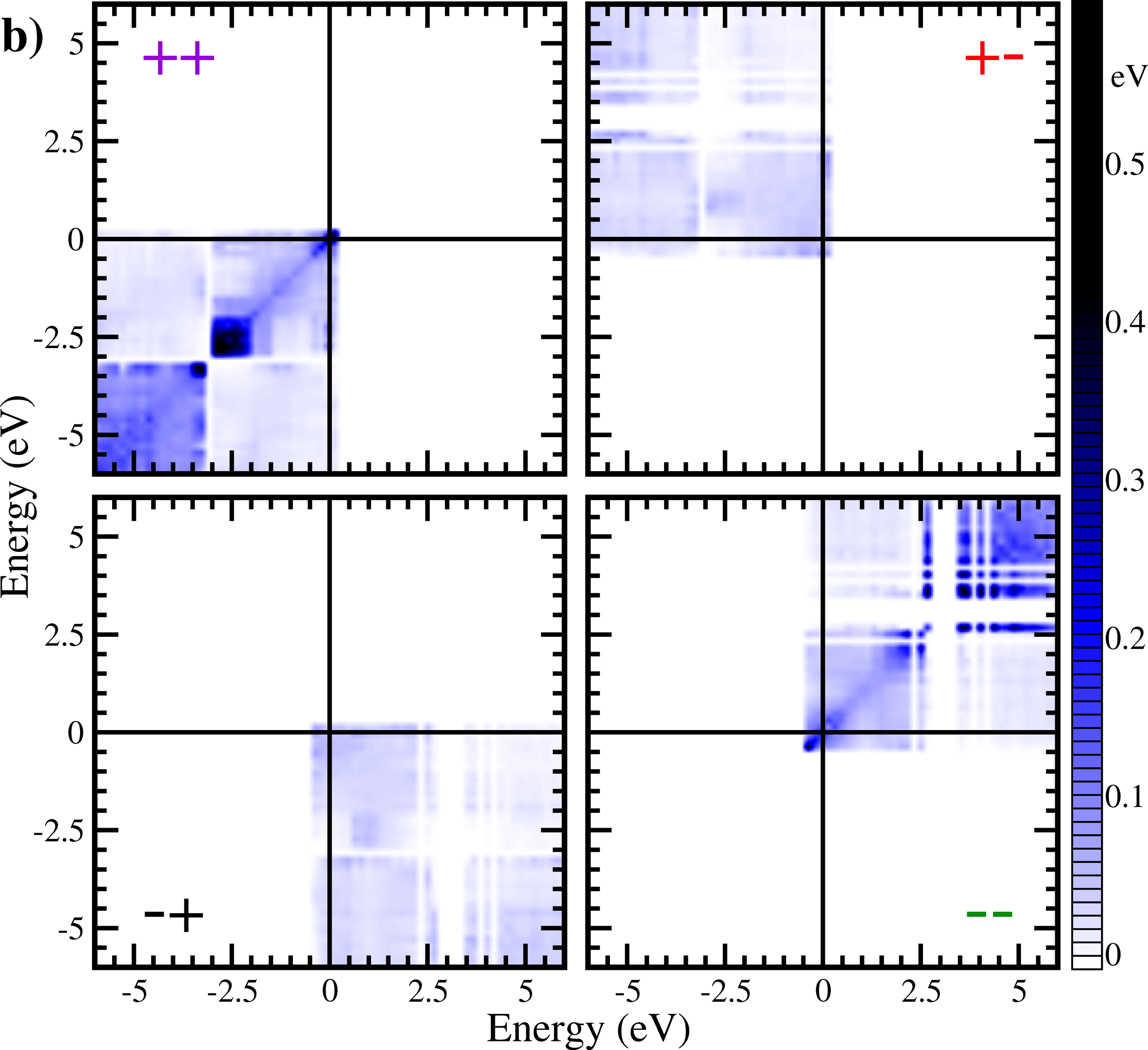}
\end{minipage}
\begin{minipage}{1.0\columnwidth}
\hspace{0.032\columnwidth}\includegraphics[width=0.79\columnwidth]{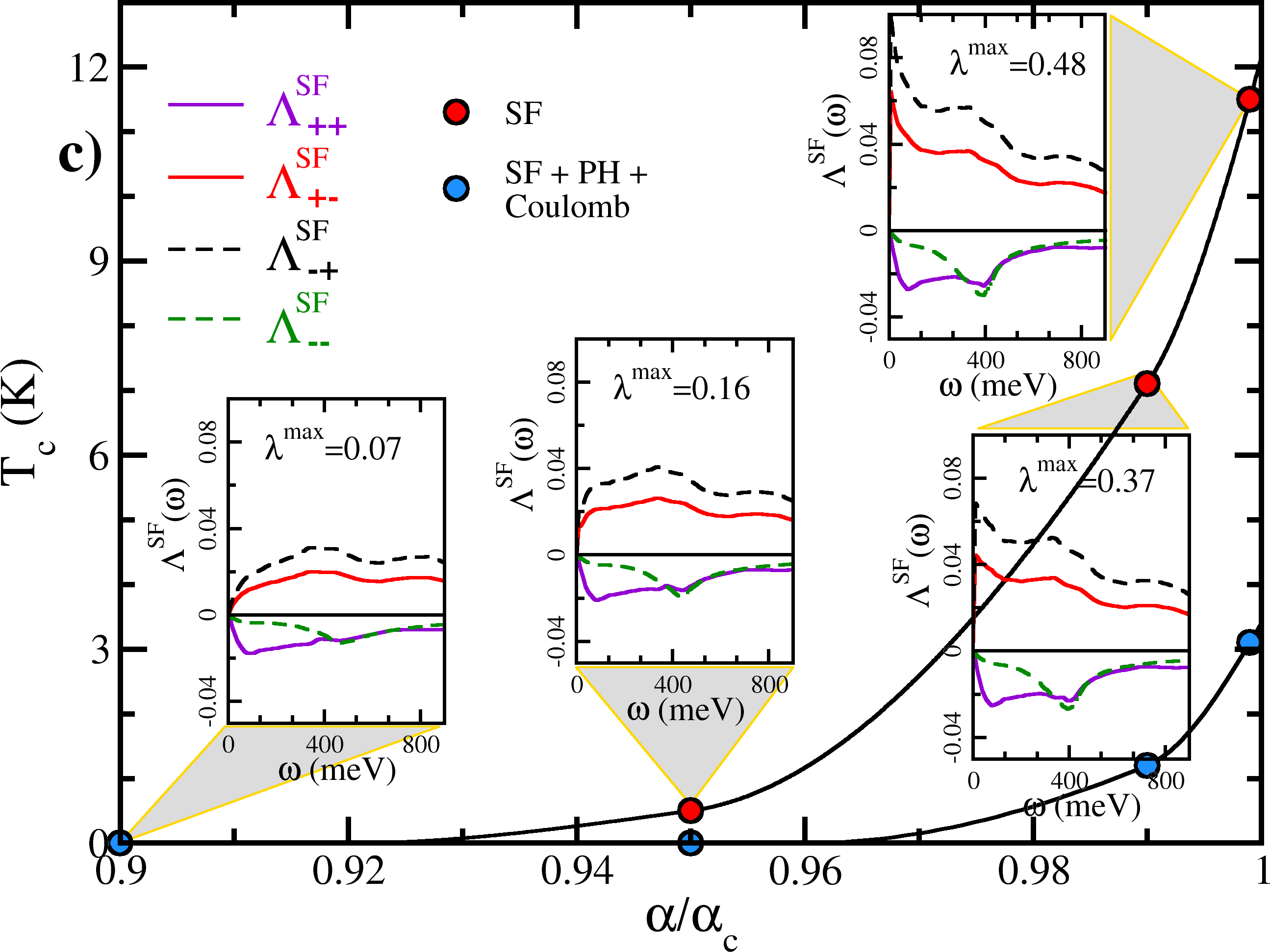}
\end{minipage}
\caption{(color online) Two bands resolved superconducting pairing functions evaluated at the theoretically optimized \zse. 
\textbf{a)} electron-phonon Eliashberg function (PH) ; 
\textbf{b)} Screened Coulomb matrix elements averaged on iso-energy surfaces (Coulomb); 
\textbf{c)} Spin fluctuation Eliashberg function (SF) and SCDFT critical temperatures (\tc) as a function of  the scaling parameter $\alpha$.}\label{fig:pairing}
\end{centering}
\end{figure}

Calculations show
\footnote{ Phonons and el-ph matrix elements have been calculated within density
functional perturbation theory\citep{RevModPhys.73.515} implemented
in quantum espresso \citep{QE-2009}. A $10\times10\times6$ and $6\times6\times4$
grid is used respectively for electronic and phononic sampling of
the Brillouin zone. Core states are threated in the ultrasoft pseudopotential
approximation. The planewave(charge) expansion is cut at 40(400) Ry. 
The electronic response is also computed by means of liner response
DFT, as implemented in the Korringa-Kohn-Rostoker approximation\citep{PhysRevLett.102.247206}.
The Brillouin zone is sampled with a grid of $16\times16\times4$ points 
in order to properly sample the peak at $q_{\mathrm{M}}$. Since the 
excitations are at low frequency a logarithmic grid in $\omega$-space
with a cut of 1eV is used.}
the el-ph coupling (see Fig.~\ref{fig:pairing}a) to be very small, in agreement with previous works\citep{PhysRevLett.101.026403,arXiv.1306.2925}.
With the integrated coupling $\lambda$ 
being less that 0.1, \tc\ would be exponentially small
if this was the only pairing channel. The Coulomb
pairing (reported Fig.~\ref{fig:pairing}b) is as expected, 
diagonally dominated (++ and -- components). Within the static approximation, this cannot
lead to any pairing, therefore no superconductivity can be sustained
at any temperature by the combined effect of Coulomb forces and phonons.

The only possible superconductivity pairing must lay in the SF term.
By the reasons discussed above this term depends critically
on the parameter $\alpha$. This can be appreciated with the help of Fig.~\ref{fig:pairing}c). 
Here $\Lambda_{ij}^{{\scriptscriptstyle \text{SF}}}$  is shown as a function of the $\alpha/\alpha_{c}$ ratio. 
At $\alpha/\alpha_{c}=$0.9 the pairing is negligible. The maximum eigenvalue\cite{PhysRevLett.3.552} 
of the $\lambda_{ij}$ matrix being $\lambda^{\rm max}=$0.07, comparable with the phononic pairing.
In the limit of $\alpha\to\alpha_c$, $\lambda^{\rm max}$ rises up to 0.48. In spite of the fact that 
the susceptibility will diverge at $\alpha=\alpha_c$ and ${\bf q}=M$ 
this value is essentially the highest limit that can be reached after 
integrating over the Brillouin zone. 
In combination with the very high SF characteristic frequencies, such a coupling 
leads to a sizable critical temperature and, since (in this limit) SF are dominated by 
off diagonal components, to the expected $s\pm$ symmetry\cite{nature.5.141}.

If only SF coupling is considered a \tc\ as high as 11~K is found.
However this critical temperature is reduced both by the inclusion
of Coulomb terms (11~K $\to$ 4~K) and phonons
\footnote{The fact that phonons reduce the \tc\ in this material is a direct
consequence of the stronger interband than the intraband coupling $\lambda_{ij}$ ($\sim0.03$
intraband and $\sim0.05$ interband), effectively acting against the
$s\pm$ ordering.} (11~K $\to$ 10~K), leading to an estimated maximum \tc\ of 3~K.

This estimation for \tc\ is in reassonable agreement with the experimentally observed
8~K\citep{Hsu23092008}. This is an important success
of the theory,  showing that the SF are indeed the origin of the 
superconductivity in FeSe and giving \tc\ of the same order of magnitude as the 
experimental one.
{\color{darkgreen2}
However, the inability of the standard DFT to describe the
ground state of FeSe enforced us to introduce parameter $\alpha$ that 
we cannot determine from the first-principles. The estimation of \tc\
also depends sensitively on the underlying electronic system,  
both via the KS electronic structure (computed at the Se Whyckoff position z$_{{\rm Se}}$=0.25%
\footnote{This value is obtained by structural relaxation on the magnetic unit
}) and the calculated spin susceptibility.}

\begin{figure}
\begin{centering}
\includegraphics[width=1\columnwidth]{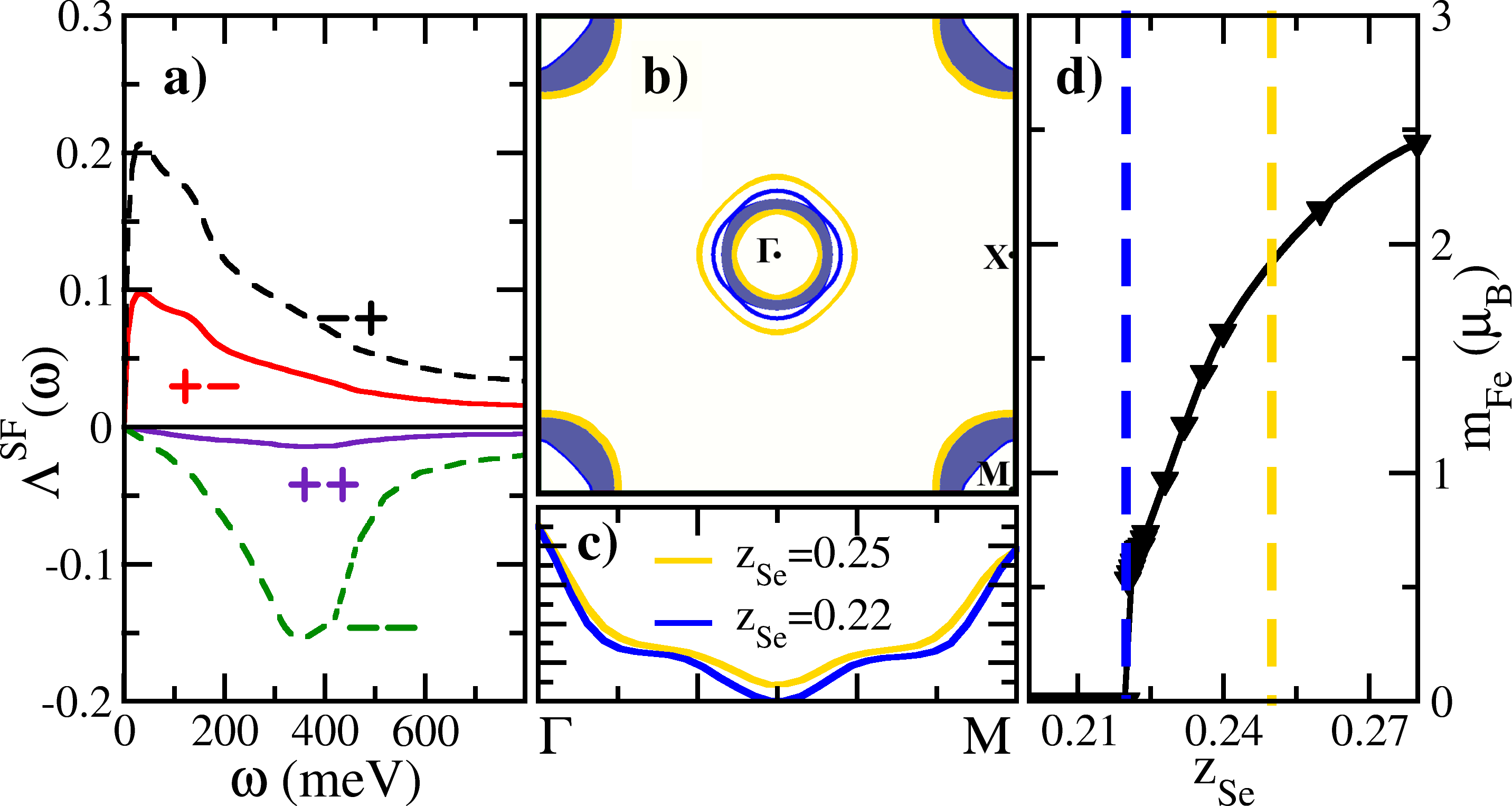} \caption{Effect of the 
variation of \zse\ position. \textbf{a)} Spin fluctuation spectral
function $\Lambda^{{\scriptscriptstyle \text{SF}}}$ computed at \zse=0.22,
leading to $\lambda^{{\rm max}}=1.3$. \textbf{b)} projection of the
Fermi surface (FS) on the xy-plane. Blue (dark) computed at \zse=0.22,
and Yellow (gray) at \zse=0.25. \textbf{c)} Nesting function (normalized)
along the $\Gamma-M$ line. \textbf{d)} magnetic moment at the Iron
site as a function of the \zse\ parameter. The values used in this
work are marked by vertical dashed lines. }

\par\end{centering}

\centering{}\label{fig:zse} 
\end{figure}

{\color{darkgreen2}
To enforce our findings we perform the SCDFT calculations 
using an alternative way to overcome the problem of the description of the 
magnetic ground state of FeSe.
As shown in Fig.~\ref{fig:zse}d
the magnetic properties of FeSe strongly depend on the Wyckoff  
positions of the Se atoms, z$_{{\rm Se}}$.
The calculations show that the system is 
 magnetic for \zse>0.22.  
We fix \zse=0.22 in the paramagnetic region close to the transition to the AFM state. 
 The proximity to the phase boundary results in intense paramagnon fluctuations.  
 This leads in our theory to large SF pairing functions (see Fig.~\ref{fig:zse}a)
 and to an estimated \tc\ of 24~K (by including all three pairing channels, the SF-only calculation gives \tc\ of 
32K).
The detailed analysis shows that the large difference in the value of \tc\
obtained in our two approachs is not due to the form of the spin susceptibility
but due to a different electronic structure near the Fermi energy. This second case provides a sharper nesting for the SF pairing (see Fig.~\ref{fig:zse}c) and leads to an increased interband pairing.}

This large sensitivity of the predicted \tc\ to the lattice
properties is consistent with experimental observations
\citep{JACS.130.3296,JPSJ.78.023701,NatureMat.8.630,PhysRevLett.102.177005,PhysRevB.80.014506},
in particular with the observed correlation of \tc\ with the anion
position\citep{SUST_Anion_h_dependence}.

\begin{figure}
\begin{centering}
\begin{minipage}[c]{0.5\textwidth}
 \includegraphics[width=1\columnwidth]{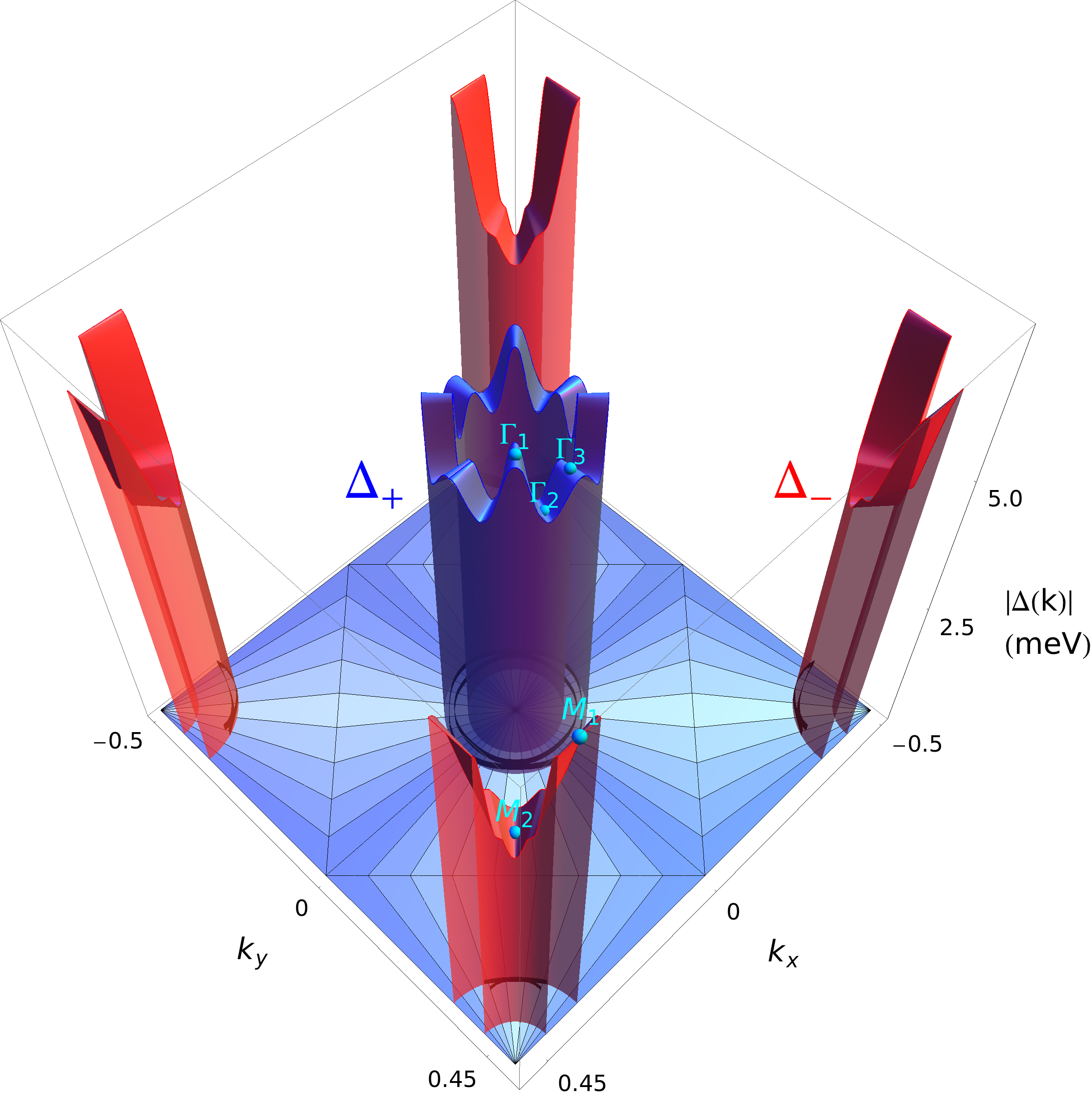}\vspace{-9.0cm}
 
\begin{minipage}[c]{1\textwidth}
 \hspace{1cm} \includegraphics[width=0.85\columnwidth]{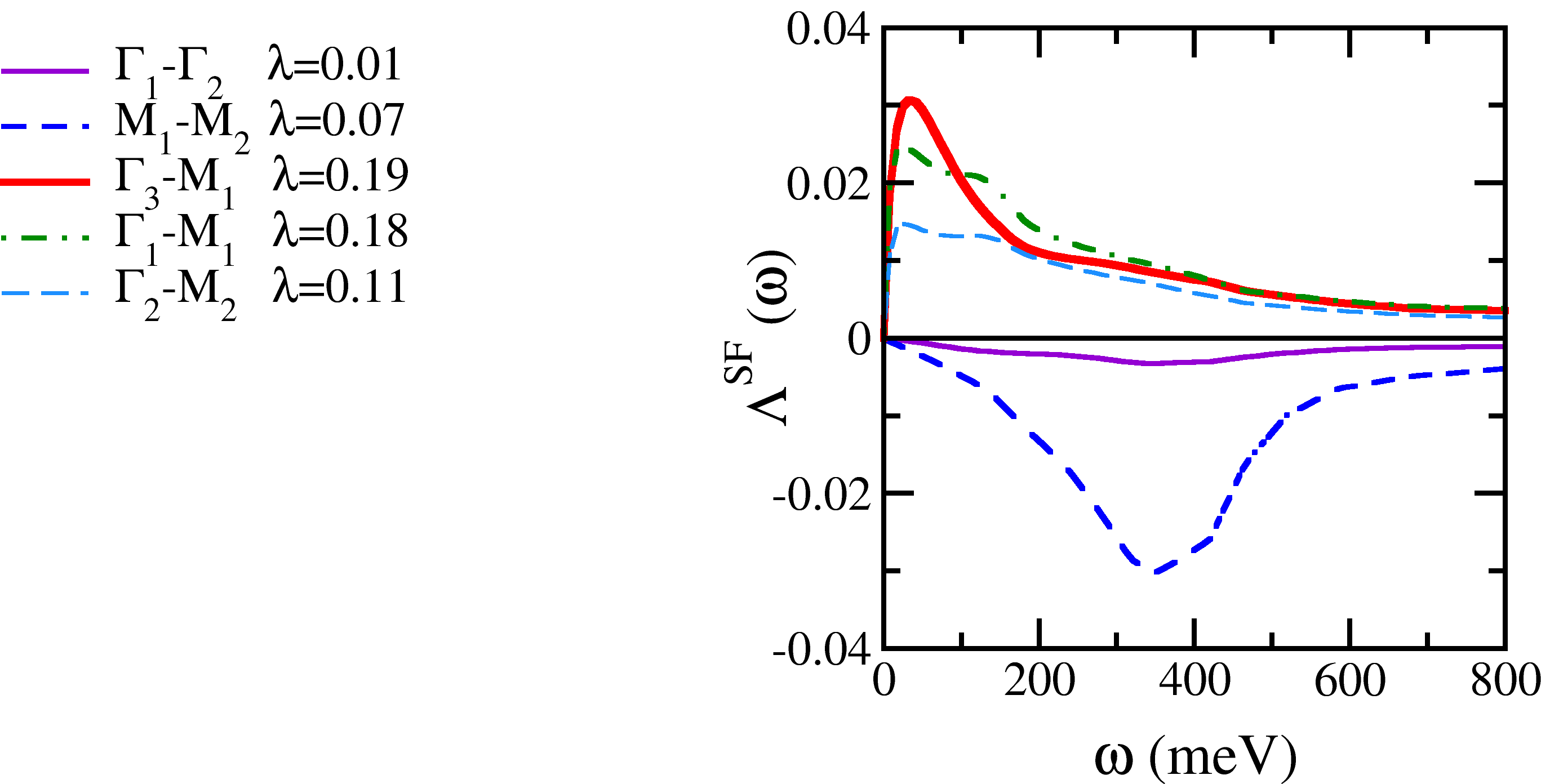} 
\end{minipage}\vspace{6.5cm}
\end{minipage}
\par\end{centering}

\centering{}\caption{Many body gap function in $k-$resolved calculation at $T=0$ and
$\omega=0$. The black lines are the FS at $k_{z}=-\frac{\pi}{2c}$.
\label{fig:Gap_of_k}}
\end{figure}

We conclude the paper with an investigation of the gap function in
 $k$-space, going beyond the two-band isotropic approximation used
so far (Eq.~\ref{eq:isotropeApp}). We divide the $xy$-plane
of the Brillouin zone into sectors as shown in Fig.~\ref{fig:Gap_of_k},
and compute the superconducting gap corresponding to each sub-band.
The use of this more accurate approach does not significantly
affect the value of the critical temperature (\tc\ increases by a few percent),
it leads to a significantly modulated gap function within the $s_{\pm}$
symmetry. The large dip in the $\Delta_{-}$ is related to the CB
fluctuation $(M_{1}-M_{2})$ in this band and is consequently not
seen in $\Delta_{+}$ $(\Gamma_{1}-\Gamma_{2})$. The smaller oscillations
are related to intraband scattering, where the interaction is smaller
between two minima of the gap $(\Gamma_{2}-M_{2})$ and larger between
two maxima $(\Gamma_{1,3}=M_{2})$.

In this work we report the first application of the SCDFT theory
taking into account the SF. In particular we consider
the case of FeSe. We demonstrate that the SF are indeed the physical
mechanism leading to the formation of the Cooper pairs and superconductivity
of the system. We demonstrate that the pairing symmetry is of the .... type.
To overcome the problem of the standard DFT theory with the description of the
FeSe ground state we adopt to different approachs that allow us to reveal 
strong sensitivity to the details of the electronic structure.
The estimated \tc\ varies between 3~K and 24~K that is in reasonable correlation 
with experimental value of 8~K.
We believe that as soon as the problem of the DFT description of the magnetic 
ground state of FeSe will be solved the suggested machinery will provide 
a complitely ab-initio estimation of the superconducting \tc\ in this compound
and will open an avenue for the first-principle design of the systems with 
high-\tc\ supeconductivity.

\bibliographystyle{apsrev4-1}
\bibliography{references}
\end{document}